\documentclass[amsmath,amssymb,reprint,superscriptaddress,pra,aps]{revtex4-2}
\usepackage[english]{babel}
\usepackage[utf8]{inputenc}
\usepackage{mathtools}
\usepackage{physics}
\usepackage{xcolor}
\usepackage{graphicx}
\usepackage[T1]{fontenc}
\usepackage{lipsum}
\usepackage{lineno}
\usepackage[colorlinks=true, allcolors=blue]{hyperref}
\usepackage{float}
\usepackage{orcidlink}
\addto\captionsenglish{}

\begin{document}
\title{Role of seeding in the generation of polarization squeezed light by atomic Kerr medium}

\author{Eduardo C. Lima\,\orcidlink{https://orcid.org/0000-0003-0239-0340}}
    \affiliation{Centro de Ci\^{e}ncias Naturais e Humanas, Universidade Federal do ABC-UFABC, Santo Andr\'{e} 09210-580, Brazil}

\author{Breno Marques\,\orcidlink{https://orcid.org/0000-0002-2560-7162}}    
    \affiliation{Centro de Ci\^{e}ncias Naturais e Humanas, Universidade Federal do ABC-UFABC, Santo Andr\'{e} 09210-580, Brazil}

\author{Marcelo Martinelli\,\orcidlink{https://orcid.org/0000-0002-6504-7040}}
    \affiliation{Instituto de F\'{i}sica, Universidade de S\~{a}o Paulo-USP, P O Box 66318,S\~{a}o Paulo 05315-970, Brazil}
    
\author{Luciano S. Cruz\,\orcidlink{https://orcid.org/0000-0003-0593-7495}}
    \email{luciano.cruz@ufabc.edu.br}
     \affiliation{Centro de Ci\^{e}ncias Naturais e Humanas, Universidade Federal do ABC-UFABC, Santo Andr\'{e} 09210-580, Brazil}

\date{\today} 

\begin{abstract}

Quantum state production and characterization are fundamental elements for many quantum technological applications. In this paper, we studied the generation of polarization quantum states by interacting light with a Kerr medium and the dependency of the outcome on orthogonal polarization seeding. Starting from coherent states produced by Ti:Sapphire laser, interaction with a $^{87}$Rb warm vapor cell led to noise compression of $-5.2\pm 0.5$ dB ($6.4\pm 0.6$ dB after correction of the detection quantum efficiency). Experimental characterization of the effect of an orthogonal polarization light seed on squeezing is shown to agree with the theoretical model.
 
\end{abstract}


\maketitle

\section{Introduction}

Quantum information provides new tools for understanding and overcoming many modern problems and technological challenges \cite{Dumke2016, Acin2018}. New quantum-based technologies receive significant attention from academics to the market, given all possible outcomes, such as quantum sensing \cite{yu2020}, quantum secure communication \cite{Qi2021}, and quantum information processing \cite{Nielsen2012}. In general, these technologies are based on intrinsic quantum proprieties such as superposition and entanglement of states \cite{bell_aspect2004}. The importance of engineering \cite{Puri2017}, production \cite{choi2023}, and measurement of quantum states \cite{Nape2021} cannot be underestimated.

Light is an interesting carrier for quantum information since it normally has low losses by interaction with the environment \cite{Polzikbook}. On the other hand, the selection of an adequate medium is necessary for information processing \cite{boyd}.
A usual approach to describe quantum features of light in a given mode uses continuous variables, by the mean value of the amplitude fluctuations, that can be associated with quadratures related to intensity and phase \cite{gerry2023}. A useful tool for information processing,  squeezed states present a reduced fluctuation in one of the quadratures when compared to corresponding coherent states. Squeezed quantum states have been studied since the 1980s \cite{walls1983}, with a large range of applications \cite{Andersen2016}, including quantum sensing \cite{Lawrie2019, Frascella2021}, quantum metrology \cite{Zhang2014}, continuous variable quantum computation \cite{jensen2011}, and quantum information \cite{Cardoso2021}. An outstanding feature is their current use to obtain the emblematic results by the effort on LIGO (Laser Interferometer Gravitational-Wave Observatory) experiments \cite{LIGO2011}.

One interesting way to produce squeezed states is by interacting light in a coherent state with a medium with second-order $\chi^{(2)}$ \cite{villar2005,furusawa1998,jing2006,pereira1992} or third-order $\chi^{(3)}$ nonlinearities \cite{ferreira2020,zhang2020,montana2020}. Since the first experiments on squeezed states \cite{slusher1985}, neutral atoms have been successfully used as these media. They were used in many applications such as memories \cite{Borba2017}, magnetometers \cite{Florian2010},  precise measurements \cite{Pezze2018}, and quantum sensing \cite{Lawrie2019} in general. Squeezed quantum states generated by atomic medium led to many experimental examples \cite{bowen2002,barreiro2011,horrom} and theoretical works \cite{chirkin2015polarization}. There is a broad list of useful properties for many applications, discovered over the years \cite{Polzik2010}.

Squeezed states can be translated to the measurement of the polarization of light. The classical polarization state of light can be measured directly on a simple detector, and the possible set of measurements can be reached by combinations of wave plates followed by a polarizing beam splitter (PBS). The polarization of a quantum state of light can be manipulated in the same way, and can be described with the help of Stokes Operators \cite{luis2016}.  For an intense field in a given polarization, the fluctuations of the outcomes of Stokes Operators are driven by the noise of the orthogonal vacuum mode. Among the techniques for the production of polarization squeezing, polarization self-rotation in a Kerr medium has proven its versatility, by the generation of squeezing with nearly degenerate modes for pump and generated fields \cite{Matsko, Novikova2009, Ries2003, Mikhailov2008}.

 The optical Kerr effect is a special case of four-wave mixing (a nonlinear process with four coupled modes). In the degenerate case, the electric-field components have the same frequency which can result in the Kerr effect as self-phase modulation or cross-phase modulation  \cite{chiao2008}, both characterized by an effective dependency on the medium refractive index with the intensity of light. The interaction with the Kerr medium with an elliptically polarized light causes the medium to become circularly birefringent: the two circular components of different intensities will propagate with different phase velocities \cite{Polzikbook}.  This effect can be produced in any Kerr material, but atomic vapors, which have strong nonlinearities near resonance without significant absorption, are especially good candidates for impressive results \cite{Glorieux2023}.  Atomic warm vapor has been a good medium to manifest the Kerr effect and was used for several applications such as squeezing light generation \cite{Korolkova2002}, quantum memories \cite{alexander2009, Heshami2016}, atomic clocks \cite{leroux2010, Louchet-Chauvet2010}, and magnetometers \cite{Horrom2012, Otterstrom2014}.

 Due to all the presented aspects, polarized light propagating through a Kerr medium is a great candidate to produce squeezed light. In fact, from previous attempts~\cite{barreiro2011}, $2.9$ dB of squeezing was observed for a given Stokes Operator. In this paper, we obtained an improved measured squeezing of $5.2 $ dB (value without any correction by losses) using a low-noise laser, in a setup with few components resulting in a well-controlled and robust source of squeezed states. We also showed the dependence of squeezing levels on the power of the seed field in the orthogonal polarization. As explained before \cite{chirkin2015polarization}, the presence of seeding is necessary to observe squeezing.

 Along the paper,  we revised the polarization squeezing and presented our model using a Kerr nonlinear media Hamiltonian in section~\ref{BG}. The experimental setup is described in section ~\ref{exp}, while the result of the best-squeezed state found and the deterioration of the squeezing by seed extinction is shown in section~\ref{sec:results}. Section~\ref{conc} is reserved for the conclusions.

\section{Background}
\label{BG}

To fully understand the model used in this paper, we revisited basic concepts of polarization and squeezed states.

\subsection{Squeezing and Variance}

If the commutation of two operators $\hat A$ and $\hat B$ results in other operator $i \hat C$, then the variances of $\hat A$ and $\hat B$ relate as the generalized 
uncertainty principle (Robertson's uncertainty relation \cite{Robertson})
\begin{equation}
    V_A V_B \geq \frac{1}{4}|\langle \hat C \rangle |^2, \label{eq:2.1}
\end{equation}
where $V_A = (\Delta \hat A) ^2 =  \langle \hat A \rangle ^2 - \langle \hat A ^2 \rangle$, and correspondingly for  
$\hat B$. Squeezing happens when one variance falls below the minimum value
\begin{equation}
    V_A \; \textbf{or}\; V_B < \frac{1}{2}|\langle \hat C \rangle |.\label{eq:2.2}
\end{equation}
As the variance of one operator is under the minimum value, it is said to be squeezed, while the other variance, which will be above the same value, is said to be anti-squeezed. 

\subsection{Polarization Squeezing}

Any polarized light can be fully described by the Stokes Parameters, which in quantum terms can be translated as the Stokes Operators \cite{bowen2002} 

\begin{equation}
\begin{split}
    \hat S_0 = \hat a_h ^\dagger \hat a_h + \hat a_v ^\dagger \hat a_v, &\qquad \hat S_1 = \hat a_h ^\dagger \hat a_h - \hat a_v ^\dagger \hat a_v, \\
    \hat S_2= \hat a_h ^\dagger \hat a_v + \hat a_v ^\dagger \hat a_h, &\qquad \hat S_3 = i(\hat a_v ^\dagger \hat a_h - \hat a_h ^\dagger \hat a_v), \label{eq:2.3}
\end{split}
\end{equation}
 where $\hat{a}_j$ ($\hat{a}^\dagger_j$) is the annihilation (excitation) operator of the polarization mode, with $j=\{h,v\}$ describing the horizontal and vertical orientations. Stokes operators are related to number operators, $\hat n_j = \hat a^\dagger_j\hat a_j$, thus 
the operator 
 $\hat S_0$ is related to light intensity, $\hat S_1$ to power unbalance between horizontal and vertical polarization, $\hat S_2$ to power unbalance between linear diagonal polarizations (linear superposition of both horizontal and vertical modes), and $\hat S_3$ to power unbalance involving circular polarizations.

The Stokes Operators lie under the $\mathfrak{su(2)}$
 \;algebra, having commutation relations as $[\hat S_i, \hat S_j] = 2i \epsilon_{ijk} \hat S_k$ with $i,j,k = \{1,2,3\}$, such as angular momentum. This implies three different uncertainty relations:
\begin{equation}
    V_1 V_2 \geq | \langle \hat S_3 \rangle | ^2,  \quad
    V_2 V_3 \geq | \langle \hat S_1 \rangle |^2, \quad
    V_3 V_1 \geq | \langle \hat S_2 \rangle |^2 . \label{eq:2.4}
\end{equation}
Those are different from usual quadrature relations, in which the right side is given by a complex constant.
As a direct consequence, more than one operator can be squeezed at the same time while the third one must be anti-squeezed \cite{bowen2002}. For cases where all the modes of light are in coherent states, considered the closest approach to classical light, all variances reduce to the same $V_0 = V_j = V_{coh} = \Bar{n}$,  the mean photon flux. 

Polarized Squeezed Light can be obtained via the interaction of light with a Kerr medium in the right conditions.
One can describe the interaction of light with a Kerr non-linear medium by the Hamiltonian \cite{chirkin2015polarization,chiao2008}
\begin{equation}
    \hat H_I = \frac{\hbar}{2} \qty(\gamma_h \hat a_h ^{\dagger 2} \hat a_h ^2 + \gamma_v \hat a_v ^{\dagger2} \hat a_v^2 + 2\gamma \hat a_h ^\dagger \hat a_h \hat a_v ^\dagger \hat a_v). \label{eq:2.5}
\end{equation}
Where $\gamma_j$ and $\gamma$ are proportional to $\chi^{(3)}$ and $\gamma_i$ represents the polarization self-modulation, while $\gamma$ is the polarization cross-modulation term. 
The system evolution can be described by the Heisenberg equation of motion:
\begin{equation}
     \dv{\hat a_j}{t} =  \frac{1}{i\hbar} [\hat a_j, \hat H_I] \implies \left\{ 
     \begin{split}
         i\dv{\hat a_h}{t} &=\qty(\gamma_h \hat n_h + \gamma \hat n_v) \hat a_h, \\
         i\dv{\hat a_v}{t} &= \qty(\gamma_v \hat n_v + \gamma \hat n_h) \hat a_v.
     \end{split}
     \right.
    \label{eq:2.6}
\end{equation}
One can verify that the mean number operator $\Bar{n}_j$ is invariant. 
Hence, the result is simplified as 
\begin{equation}
    \begin{split}
        \hat a_h (t) &= \exp\qty[-i t (\gamma_h \hat n_h + \gamma \hat n_v)] \hat a_h (0), \\
        \hat a_v (t) &= \exp\qty[-i t (\gamma_v \hat n_v + \gamma \hat n_h)] \hat a_v (0). \label{eq:2.7}
    \end{split}
\end{equation}
As the system's time dependence is on the operators, mean values are taken on the initial state, a coherent state $\ket{\alpha_h,\alpha_v}$. The application of those operators is described by $\hat a_j  \ket{\alpha_j} = \alpha_j \ket{\alpha_j}$, where $|\alpha_j|^2=\bar n_j$ relates to the mean photon number of each polarization.   
Using ${\gamma_j,\gamma}\ll 1$, we can calculate the variance of Stokes Operators, where the squeezing can occur in $V_2$ or $V_3$, while $V_0$ and $V_1$ stay coherent \cite{chirkin2015polarization}:
\begin{equation}
    \begin{split}
        V_2 &= \Bar{n}_h + \Bar{n}_v - 2(\gamma_h - \gamma_v)\Bar{n}_h \Bar{n}_v \sin(2\phi),\\
        V_3 &= \Bar{n}_h + \Bar{n}_v + 2(\gamma_h - \gamma_v)\Bar{n}_h \Bar{n}_v \sin(2\phi),\label{eq:2.8}
    \end{split}
\end{equation}
where
\begin{equation}
            \phi = \arg(\alpha_h ^* \alpha_v) + \Bar{n}_h \sin(\gamma_h - \gamma) - \Bar{n}_v \sin(\gamma_v - \gamma).\label{eq:2.8phase}
\end{equation}
Since the argument $\phi$ can be optimized by the relative phase between the orthogonal components, the maximum noise compression is given by
\begin{eqnarray}
   \mathcal{S}= \frac{V_j}{|\langle \hat S_1\rangle|}&=&\frac{\Bar{n}_h + \Bar{n}_v - 2|\gamma_h - \gamma_v|\Bar{n}_h \Bar{n}_v}{| \Bar{n}_h - \Bar{n}_v|} \nonumber \\
    &\simeq& 1-2|\gamma_h - \gamma_v|\Bar{n}_v
    \label{eq:2.10}
\end{eqnarray}
where the last term considers $\Bar{n}_v\ll\Bar{n}_h$ and $\mathcal{S}$ stands for squeezing.

Some interesting consequences come from these results \cite{chirkin2015polarization}. 
Anisotropy in the medium, given by $\gamma_h - \gamma_v$, comes as a necessary condition for the squeezing generation. Moreover,  it is easy to see that if there is no vertical polarization component in the incident light, $\Bar{n}_v=0$, all the variances will be coherent. So, there must be a non-zero power in the orthogonal polarization for the occurrence of polarization squeezing.

\begin{figure}
    \centering
    \includegraphics[width=0.43\textwidth]{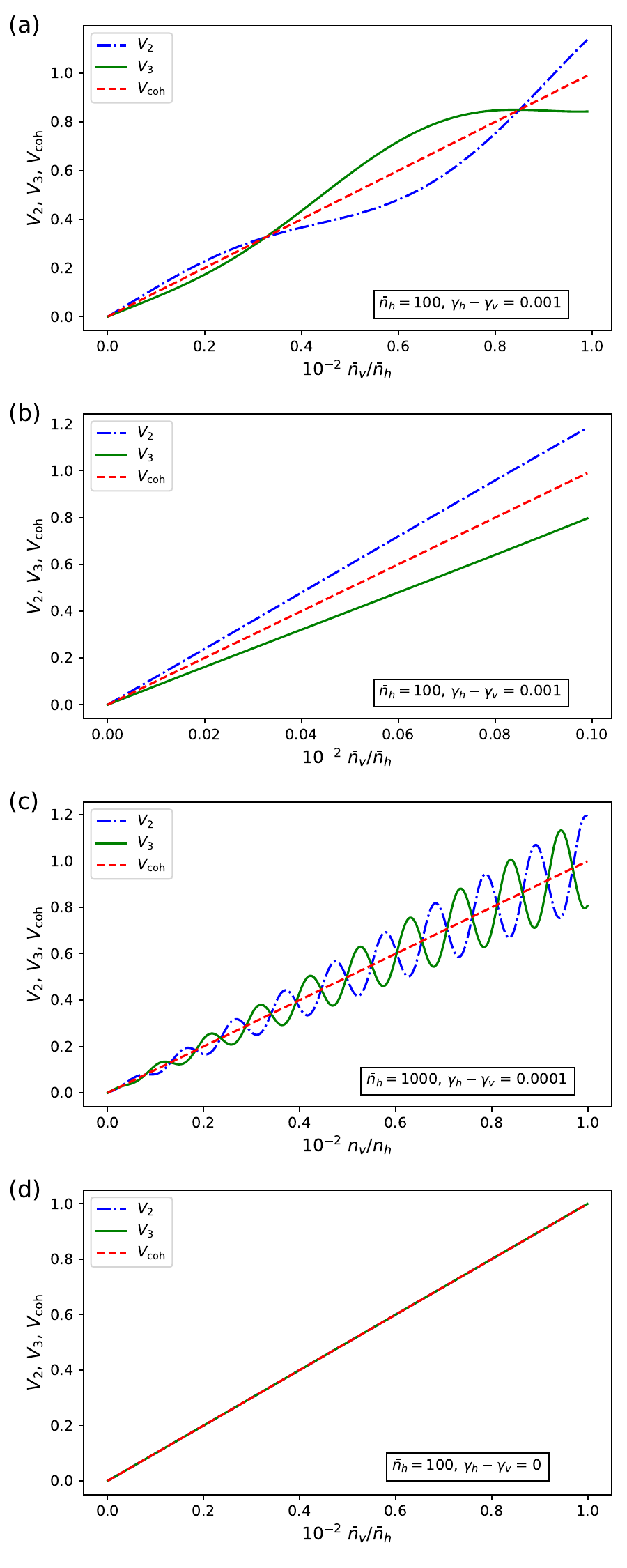}
    \caption{Theoretical variances $V_2,~V_3$, and $V_{coh}$ after passing through a Kerr medium: (a) the ratio between vertical and horizontal polarization up to $10^{-3}$ that shows the oscillatory take over the behavior, where increasing the ratio will increase the oscillation frequency; (b) the ratio is up to $10^{-4}$, where we see a linear-like function and smaller ratios will not influence further changes; (c) higher photon number on horizontal polarization, but fixed ratio showing that the higher $n_h$, the lower $\gamma_h-\gamma_v$ is needed for the system showing squeezing;
    (d) when we have an isotropic simulation $\gamma_h=\gamma_v$, no squeezing is observed. Other parameters are $\gamma_h \simeq 3\gamma$, $\alpha_h, \alpha_v$ real. }
    \label{fig:1}
\end{figure}

Simulations on distinct conditions of input photon number and anisotropies can be seen in Fig.~\ref{fig:1}. Variances of Stokes operators are compared to the expected result for a coherent state (dashed lines) as a function of the input power in the $v$ polarization, normalized by the power in the $h$ polarization. In Fig.~\ref{fig:1}(a) we can observe the evolution of the variances of $\hat S_2$ and $\hat S_3$, changing in a complementary way as described in eq.~\ref{eq:2.8} for increasing seed power. This oscillatory behavior is expected from eq.~\ref{eq:2.8phase}. If we focus on the initial evolution, as in Fig.~\ref{fig:1}(b), the nearly linear behavior is evident, and consistent with the expected evolution of a small argument in the $\sin$ function in eq.~\ref{eq:2.8}.

It is interesting to notice that increasing the pump power, while reducing the nonlinear anisotropy, will keep the compression level, if we compare Figs.~\ref{fig:1}(a, c), but will imply a faster oscillatory response. On the other hand, for an isotropic response, the outcome is reduced to variances consistent with those of a coherent state (Fig.~\ref{fig:1}(d)).

Some of these features will be explored in the experiment, that we will describe next.

\section{Experiment}
\label{exp}
\begin{figure}[ht]
    \centering
    \includegraphics[width=0.49\textwidth]{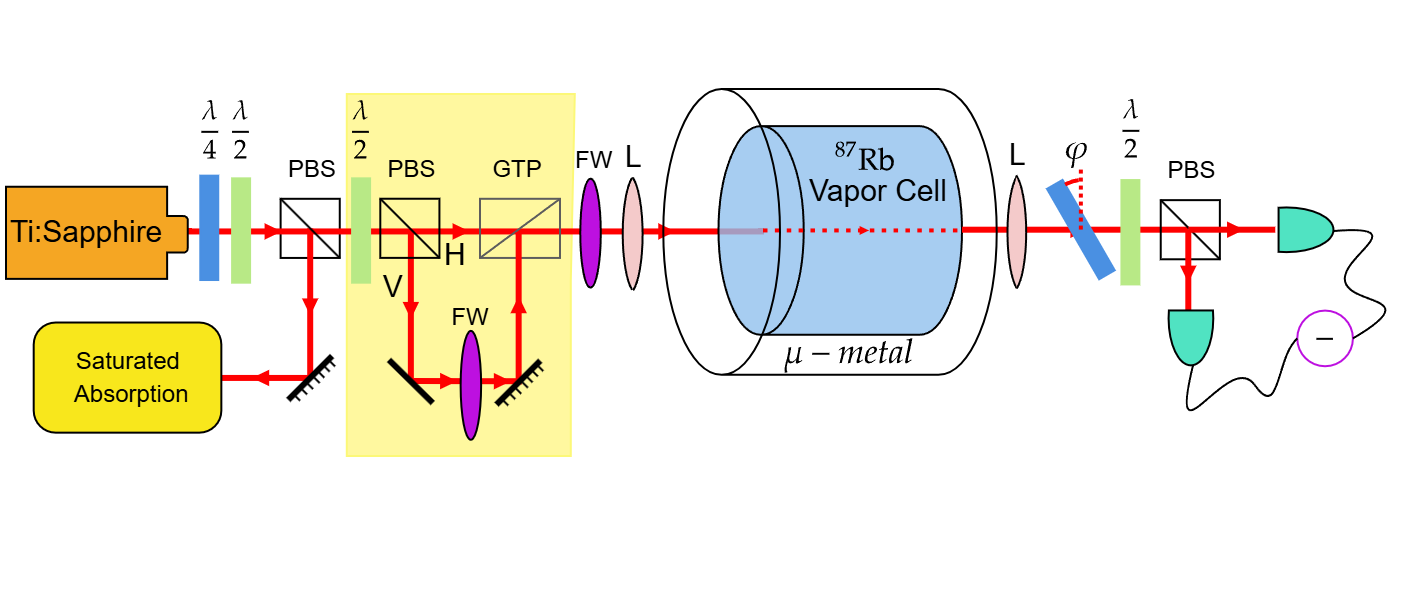}
    \caption{Experimental Setup used to measure $\hat S_\theta$. The local oscillator is co-propagating with the squeezed beam. PBS, polarizer beam splitter; $\lambda/2$, half-wave plate; $\lambda/4$, quarter-wave plate; FW, Filter wheel; L, bi-convex lens; 
    GTP, Glan-Taylor Polarizer. The light yellow (shaded) area is only present during the 
    analysis of the seed, as 
    explained in section \ref{subsec:seed}.}
    \label{fig:2}
\end{figure}

To verify the presented theory, we used an experimental setup similar to those used in \cite{barreiro2011,valente2015, horrom} (Fig.~\ref{fig:2}). A laser beam with very low noise at the analysis frequency, limited to those of a coherent state, is generated by a Ti:Sapphire laser (MBR110/Coherent), pumped by a 532 nm laser (Verdi V6/Coherent), resulting in up to 220 mW of laser light. The laser frequency can be swept around the  $D_1$ line of $^{87}$Rb, close to 795 nm. Saturated absorption is used for locking the line on the desired transition or measuring the laser frequency. The incident light on the vapor cell is almost horizontally polarized, having a tiny contribution to the vertical mode from the intrinsic transmission of the polarizing cube (PBS). The laser beam is focused, by a ($f =500$ mm) bi-convex lens, close to the center of a 75-mm-long transparent glass cell containing enhanced $^{87}$Rb (GC19075-RB/Thorlabs),  having a beam waist of $0.9$ mm close to the center.

The rubidium cell temperature is controlled by a resistance oven, ranging between $70$ and $90^\circ$ C. A triple layer $\mu$-metal shield involves the rubidium cell to minimize any external magnetic field. After passing through the cell, the laser beam is collimated by a second lens and goes through a zeroth-order quarter-wave plate $\lambda/4$ (WPQSM05-780/Thorlabs). The rotation of the wave-plate around the physical axes can be performed in a range of $\pm\; 22^\circ$.As the fast axis is aligned with the vertical component of light, and hence the slow axis with the horizontal mode, there is no change in light polarization \cite{horrom}, but the dephasing between the modes can be adjusted. 

Homodyne detection is completed by an adjustable splitting of the beam, using the polarization rotation by a half-wave plate $\lambda/2$ (WPHSM05-780/Thorlabs). The pump beam is then used as a local oscillator that can be mixed with the squeezed light, with an adjustable dephasing that is controlled by the quarter wave-plate tilt. Light is finally split by a PBS (PBS122/Thorlabs) and detected by two photocurrent detectors (photo-diodes FDS100/Thorlabs, with $81\%$ of quantum efficiency) whose photocurrent fluctuations are processed by trans-impedance amplifiers.

The resulting signals are then subtracted and measured by a spectrum analyzer (N9000A/Keysight). The data of average values and fluctuations are saved and treated with Python analysis programs. The difference fluctuation noise is compared to shot noise to define the squeezing factor of the state. 

\begin{figure}[ht]
    \centering
    \includegraphics[width=0.48\textwidth]{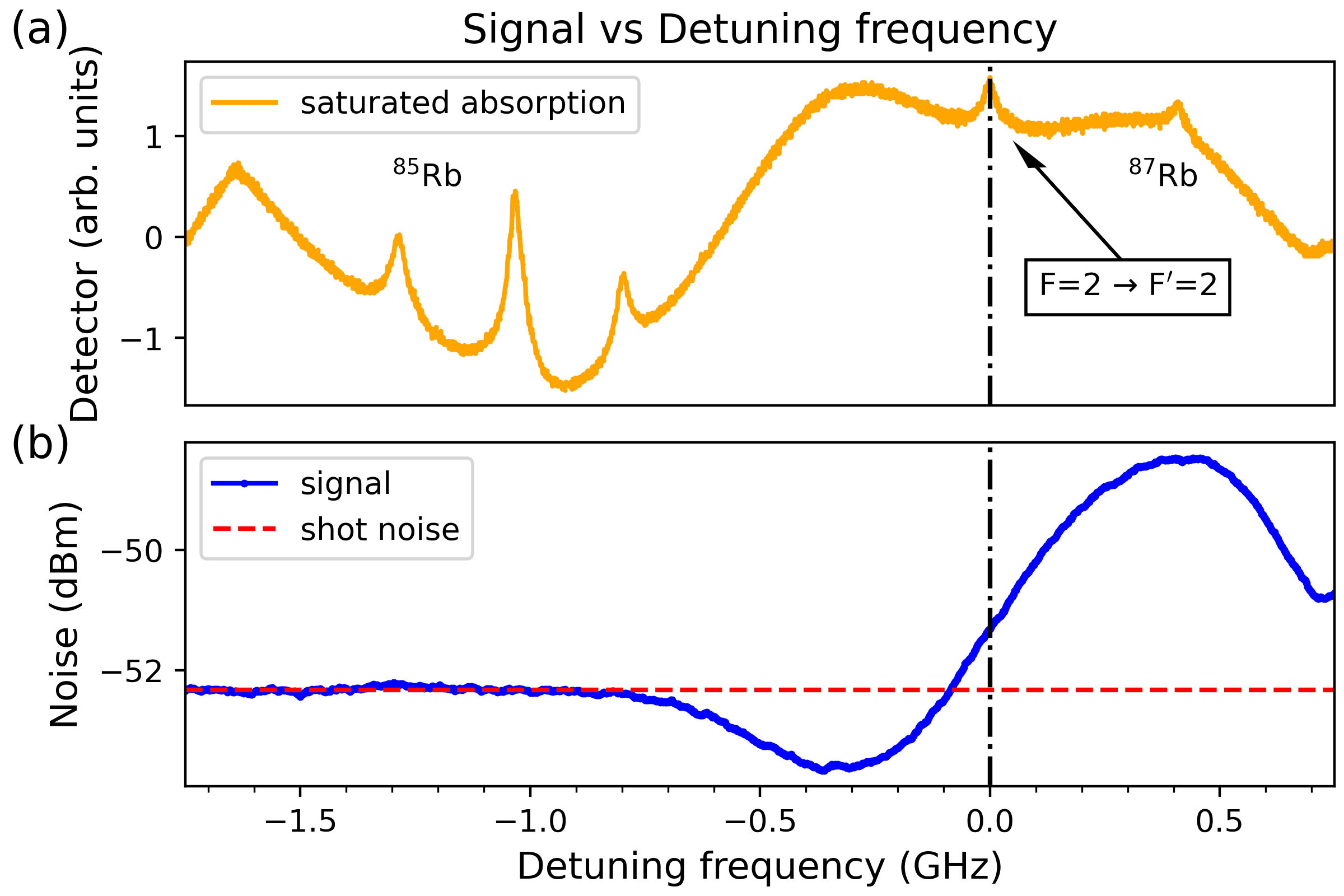}
    \caption{(a) Saturated absorption around the $F=2 \to F'=2$ transition. (b) Data acquired via Spectrum Analyzer. Sweep time $= 50$ ms. Gate length $= 50$ ms, gate delay $= 17$ ms. Average number $= 50$. Resolution Bandwidth (RBW) $= 300$ kHz, Video Bandwidth (VBW) $= 300$ Hz. Best squeezing is seen around $\Delta=-0.35$ GHz of detuning frequency.} 
    \label{fig:3}
\end{figure}

The laser frequency can be 
 swept around the $D_1$ line of $^{87}$Rb or stabilized at a given detuning.

 We chose to study the frequencies around the direct transition $F=2\to F'=2$ due to high squeezing and the possibility of acquiring the shot noise within each measurement.  One can see the saturated absorption signal in a reference cell with the natural abundance of rubidium in Fig.~\ref{fig:3}(a). The cell can serve as a frequency reference for the scale in the registration of the noise by the  Spectrum Analyzer, shown in Fig.~\ref{fig:3}(b). The electrical parameters of the Analyzer, depicted in the figure, are kept along the experiment. We verified that below $- 1$ GHz of the transition, the noise is reduced to the shot noise level, thus serving as a reference for the calibration of the spectra.

\section{Results}
\label{sec:results}

 \begin{figure}[ht]
    \centering
\includegraphics[width=0.465\textwidth]{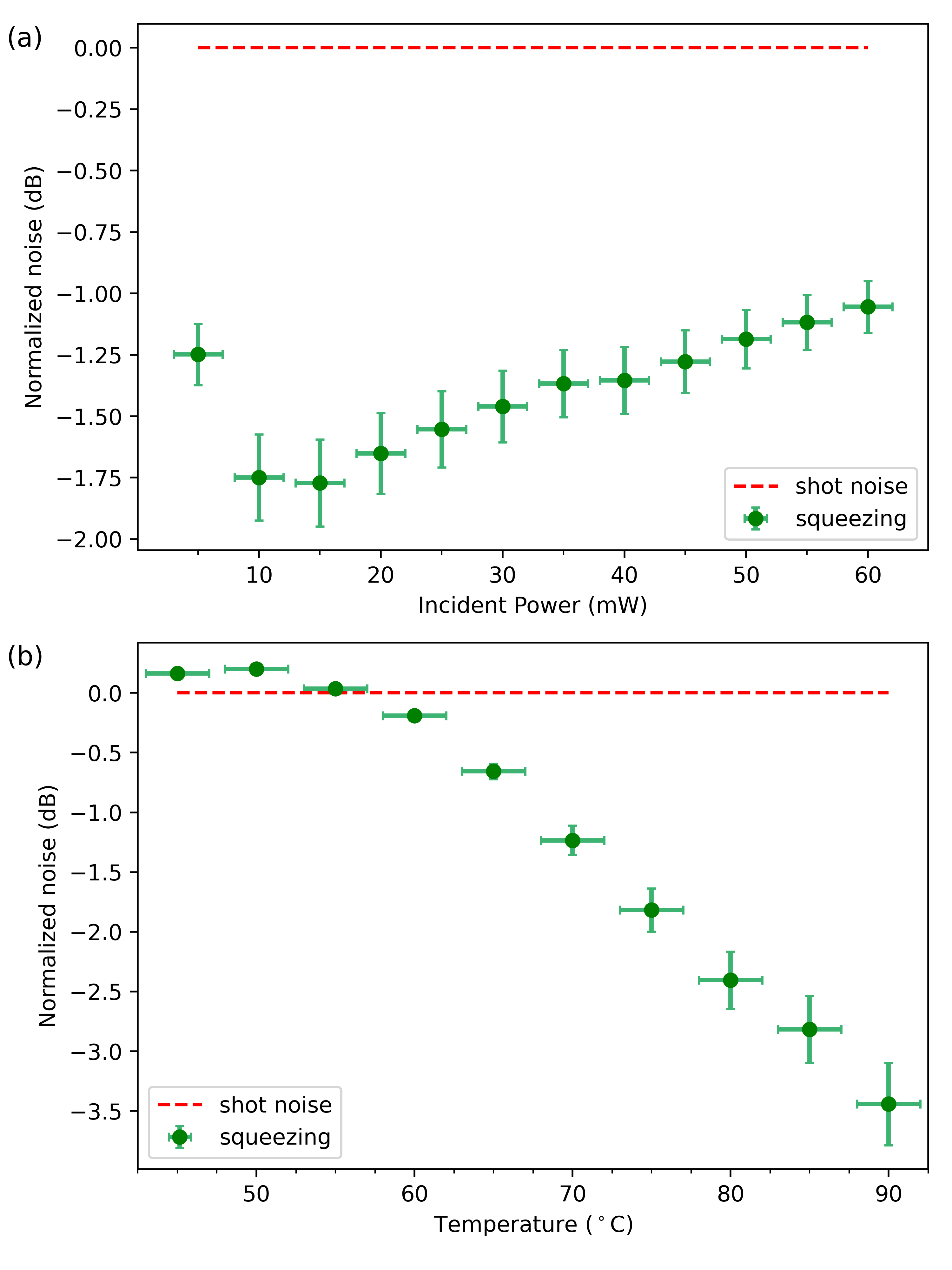}
    \caption{All measurements done with RBW = 300 kHz, VBW = 300 Hz and $\Omega = 11$ MHz. (a) Incident power optimization. Temperature kept at $72^\circ$ C. (b) Temperature optimization. Incident power at $30$ mW.}
    \label{fig:4}
\end{figure}

To find the best squeezing factor for our system, we scanned the parameters: incident power,  temperature (optical density), and tilting angle of the quarter wave-plate angle (local oscillator phase). In figure~\ref{fig:3}, we saw the behavior of our squeezing factor by adjusting the apparatus for some values of the parameters specified before. 

It is important to emphasize that the local oscillator control is made by a $\lambda/4$, just after the Rb cell, adding a phase shift between the two modes of polarization, putting the Stokes parameter into a generalized superposition
$$\hat S_{\theta} = \hat S_2 \cos \theta + \hat S_3 \sin \theta,$$ 
where $\theta$ depends on the rotation of the squeezing ellipse along the $S_2$-$S_3$ plane. 
The choice of $\lambda/4$ plate is not mandatory; one could use any birefringent material that gives different phase shifts to each polarization.
Varying the tilt angle $\varphi$, we can adjust the quadrature $\theta(\varphi)$ to scan for the best-squeezing factor.  
The best squeezing for Stokes Parameter $\hat S_2$ was consistently observed for a tilting angle of  $\varphi=\pm18^\circ$, the value that we kept for all other characterizations.

The best incident power is found to be  around $15$ mW
at temperatures between $85^\circ$C and $90^\circ$C (Fig.~\ref{fig:4}(a)). Notice that power adjustment is done by changing the wave-plate before the first PBS (Fig.~\ref{fig:2}), which implies changing both the pump and the seed. Therefore, we do not expect the curve to follow any of the models we discussed here, following \cite{chirkin2015polarization}, but rather find an optimal condition for better squeezing.

\begin{figure}
    \centering
    \includegraphics[width=0.49\textwidth]{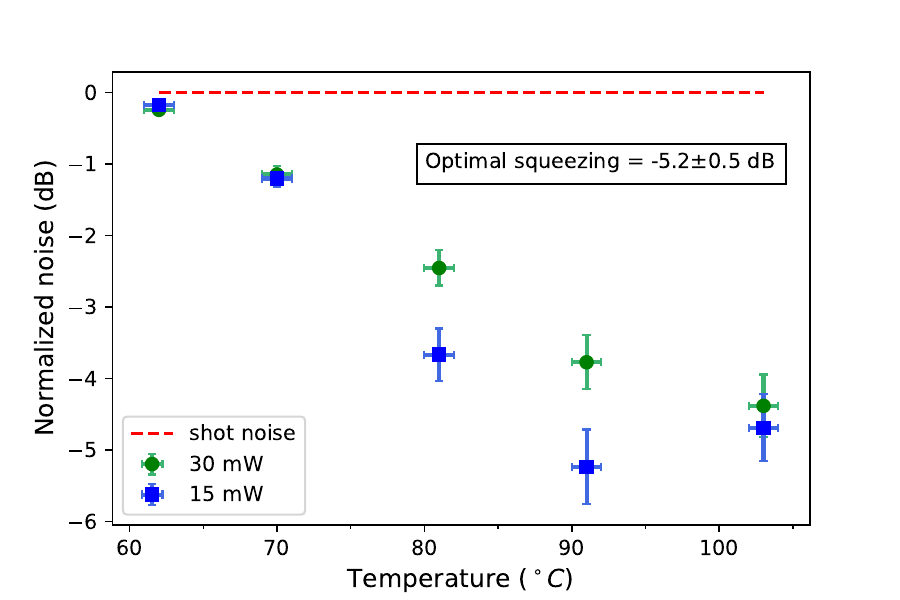}
    \caption{Optimal squeezing achieved. Tilt angle at $\varphi=-18^\circ$, analysis frequency at $\Omega=11$ MHz, detuning frequency at $\Delta = -0.35$ GHz, RBW = 300 kHz, VBW = 300 Hz.  }
    \label{fig:5}
\end{figure}

 Finally, increasing the optical depth by changing the cell temperature led to a monotonic increase in the measured noise compression (Fig.~\ref{fig:4}(b)). With these parameters evaluated, we searched for the optimal squeezing at two pump powers, finding a saturation for the compression in terms of the temperature. As shown in Fig.~\ref{fig:5}, optimal squeezing of  $\mathcal{S} = 5.2\pm 0.5$ dB is achieved at 91 $^\circ$C. If we consider the quantum efficiency of our detectors, the corrected squeezing obtained by the system is up to $6.4\pm 0.6$ dB right at the cell output. This represents a higher squeezing when compared with $2.9\pm0.1$ obtained by \cite{barreiro2011}, probably due to a higher frequency of analysis $\Omega$ and the use of a Ti:Sapphire laser, which is notoriously less noisy than a diode laser \cite{bramati}.

\subsection{Seeding}
\label{subsec:seed}

As a direct consequence of equation~\ref{eq:2.10}, squeezing will vanish if the number of photons in the vertical polarization mode is nearly zero. To study this property, we include in our setup the components indicated in the light yellow (shaded) area (see Fig.~\ref{fig:2}). 
A (PBS) separates the vertical and horizontal components, with a power ratio of 1.5 \%, adjusted by the half waveplate. These two modes are recombined in a Glan-Taylor Polarizer (GTP), chosen by its high extinction ratio, of $1:10^5$. This ensures a high purity for the transmitted horizontal polarization, with a nearly perfect extinction of the vertical component. Therefore, the vertically polarized seed contribution comes mainly from the light injected through the other port of the GTP.  The use of a neutral density wheel allows fine control of the power in the seed mode.

To assure the spatial superposition of the vertically and horizontally polarized modes after recombination in the GTP, before the measurement, we adjust the alignment maximizing the 
mode matching to more than 92\%.

Total incident power at the input of the cell is adjusted to 15 mW by 
another neutral density filter wheel (NDC-50C-2M-B/Thorlabs) just before the rubidium cell.
In this condition, the maximum squeezing factor measured was around $- 4$ dB.
 
\begin{figure}[ht]
    \centering
    \includegraphics[width=0.48\textwidth]{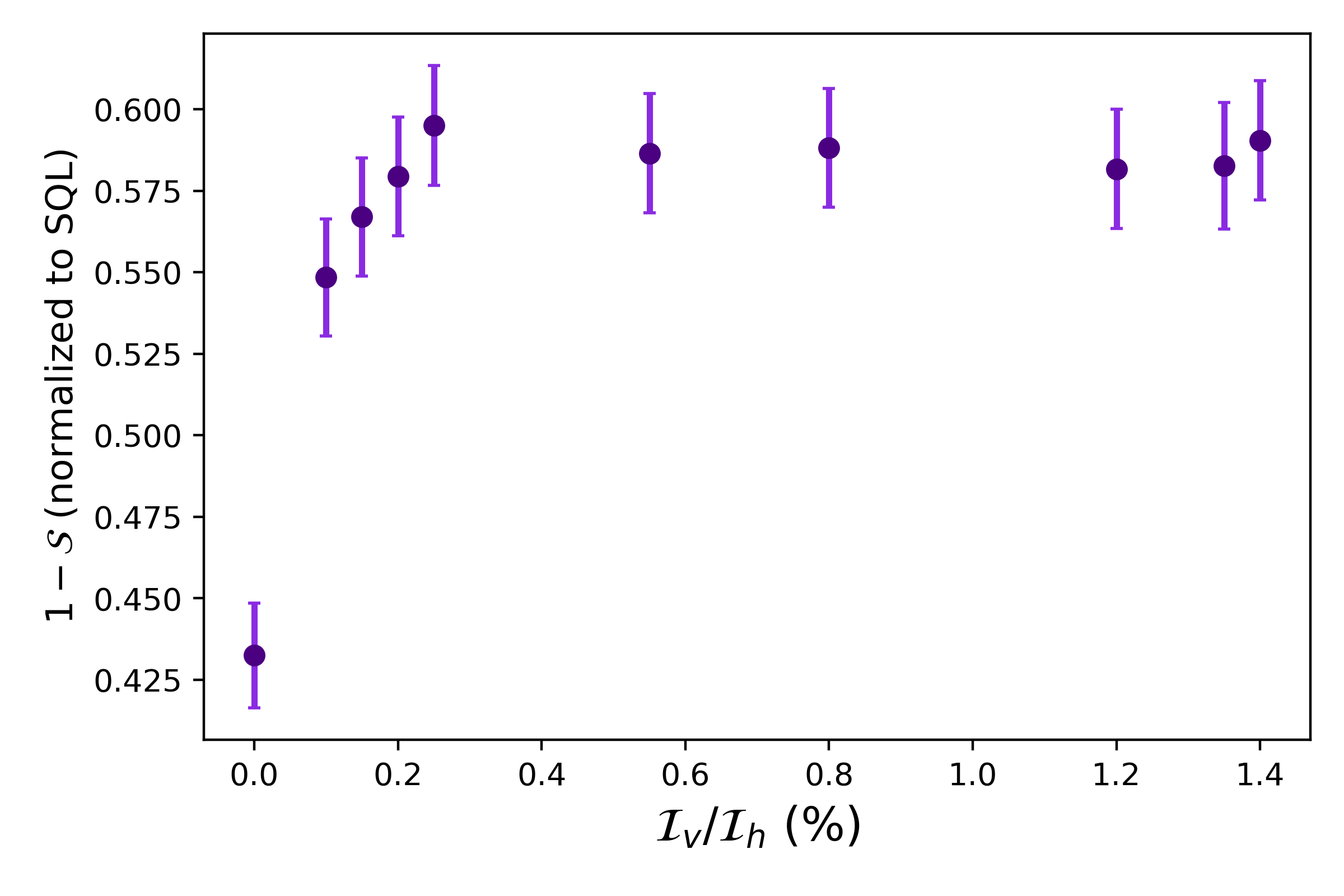}
    \caption{Normalized noise power for seed power, given by a ratio between vertically and horizontally polarized components of the incident beam interacting with the Rb atoms.}
    \label{fig:6}
\end{figure}

The squeezing dependence on the seed, in terms of the ratio of vertical and horizontal power ($\mathcal{I}_v/\mathcal{I}_h$), is shown in Fig \ref{fig:6}; the vertical axis is presented in terms of the normalized squeezing $1-\mathcal{S}$, giving the squeezing gain as the vertical seed intensity is increased (horizontal axis).

Starting from a complete blocking of the vertically polarized light, we have a monotonic increase of the noise compression, seen by the reduction of the noise (Fig. \ref{fig:6}), until compression noise reaches a plateau. The complete suppression of the incoming vertical polarization is a major challenge. Notice that even the residual transmittance of the Glan-Taylor polarizer, of less than one part in $10^5$, still results in an important squeezing value, reducing the noise to around $ 57$ \% of the standard quantum level (SQL). Nevertheless, as we increase the photon number in the seed mode, the noise compression also grows. This growth saturates at $I_v/I_h \simeq 0.25 \%$, stabilizing the noise in a plateau at $ 41 \%$ of the SQL. 
The results confirm that the seed intensity is an important factor to be taken into account in this experiment, in order to maximize the noise compression, with a possible limiting factor coming from the homogeneity of the non-linearity given by $\gamma_h-\gamma_v$ in Eq. (\ref{eq:2.10}).

\section{Conclusions}
\label{conc}
We maximized the value for a polarization-squeezing apparatus and found the optimal squeezing factor of $5.2\pm0.5 \;(6.4\pm0.6)$ dB exploring the values of incident power and Rb optical density, thus demonstrating this simple and robust setup is a reliable source of high level of squeezing. Consistency and reproducibility relied on the use of a coherent pump.

This is consistent with the high level of noise compression given by atomic vapors the in four-wave mixing process \cite{Glorieux2023}, which demonstrates these simple setups as interesting candidates for the production of nonclassical states, which may eventually compete with other techniques based on $\chi^{(2)}$ crystals \cite{Schnabel} as reliable source of quantum states for applications in quantum information.

We also investigated the role of the perpendicular polarization seed on the squeezing production. The general model \cite{chirkin2015polarization} shows a strong dependency on this seed, which is corroborated by our experimental results. The saturation was not observed in the model, which could indicate that this feature depends on the atomic electronic structure. The experiment shows the results using alkali atoms, but the theory and data can be applied in any Kerr medium system. Control of the asymmetry in the cross-phase modulation is shown as the key factor for strong compression.

\section*{Acknowledgments}

E.C.L acknowledges support from FAPESP (Grant No. 2023/03513-0). B. M. and M. M. acknowledge CNPq. All the authors acknowledge support from Brazilian National Institute of Quantum Information (CNPq-INCT-IQ Grant No. 465469/2014-0), FAPESP (Grant No. 2014/03682-8 and No. 2021/14303-1). All the authors acknowledge T. H. D. Santos for his previous work in the laboratory.

\end{document}